# Hydrodynamic simulations of galaxy formation.

# I. Dissipation and the maximum mass of galaxies.

Anne A. Thoul & David H. Weinberg

Institute for Advanced Study, Princeton, NJ 08540

E-mail: thoul@guinness.ias.edu, dhw@guinness.ias.edu

## ABSTRACT

We describe an accurate, one-dimensional, spherically symmetric, Lagrangian hydrodynamics/gravity code, designed to study the effects of radiative cooling and photo-ionization on the formation of protogalaxies. The code can treat an arbitrary number of fluid shells (representing baryons) and collisionless shells (representing cold dark matter). As a test of the code, we reproduce analytic solutions for the pulsation behavior of a polytrope and for the self-similar collapse of a spherically symmetric, cosmological perturbation. In this paper, we concentrate on the effects of radiative cooling, examining the ability of collapsing perturbations to cool within the age of the universe. In contrast to some studies based on order-of-magnitude estimates, we find that cooling arguments alone cannot explain the sharp upper cutoff observed in the galaxy luminosity function.

*Subject headings:* Galaxies:formation, Hydrodynamics, Methods: numerical

## 1. Introduction

The leading cosmological theories imply that galaxies form by the collapse of primordial density fluctuations. The gravitational evolution of collisionless matter can be followed by various dynamical approximations, or, in the strongly non-linear regime, by N-body simulations. However, gas dynamical effects such as shocks and radiative cooling must play an essential role in the formation of galaxies, since gas must cool and condense inside dark matter halos before it can form stars.

There have been two quite different approaches to this theoretical problem. One, going back to the pioneering work of Binney (1977), Silk (1977), and Rees & Ostriker (1977) (hereafter RO), uses simple analytic estimates: typically, one computes the characteristic density and virial temperature of a dark halo assuming a spherical collapse model, then asks whether gas at this density and temperature can cool within a dynamical time, or within a Hubble time. Combined with extended versions of the Press-Schechter (1974) formalism, these methods can yield detailed predictions for properties and evolution of the galaxy population (White & Rees 1978; White & Frenk 1991; Kauffman, White & Guiderdoni 1993; Cole et al. 1994). The second approach, which has become computationally practical only within the last few years, is to incorporate gas dynamics directly into three-dimensional numerical simulations (e.g. Katz & Gunn 1991; Cen & Ostriker 1992, 1993; Katz, Hernquist & Weinberg 1992; Evrard, Summers & Davis 1994; Steinmetz & Muller 1994).

In this paper and those that follow it, we will take an intermediate path, modeling the collapse of individual perturbations with a *one-dimensional*, Lagrangian, gravity/hydro code. The code evolves a mixture of gas and collisionless dark matter, elements of which are represented by concentric, spherical shells. The gas responds to gravity and pressure forces; it can be heated by adiabatic compression, by shocks, and by energy input from a photo-ionizing background, and it can cool by a variety of atomic radiative processes. The collisionless dark matter responds only to gravitational forces. While we focus in this paper on galaxy-scale collapses assuming spherical symmetry, the code is also well suited to studies of Lyman-alpha clouds, and it can easily be adapted to planar or cylindrical symmetry.

Larson (1969,1974) studied galaxy formation with spherically symmetric simulations more than 20 years ago. However, the intervening years have seen great changes in the theoretical underpinnings of galaxy formation – especially the introduction of dark matter and the development of physically motivated initial conditions – and they have seen great improvements in computational algorithms and hardware, so there is plenty of reason to revisit this approach. Our calculations include radiative cooling in the gas component and gravitational interactions with a collisionless component, and we adopt initial conditions appropriate to Gaussian random fluctuations, as might be produced by inflation in the early universe. Instead of Larson's Eulerian-grid approach, we adopt a Lagrangian representation of the gas and dark matter, which provides much higher spatial resolution in the central, high-density regions of a collapsing protogalaxy. The manyfold increase in computer power allows us to use large numbers of fluid elements and to perform faster searches in parameter space.

Our numerical approach — one-dimensional, Lagrangian hydrodynamics with radiative cooling — is similar to that used by Thomas (1988) in his models of cooling flow galaxies and by Shapiro & Struck-Marcell (1985) in their studies of "pancake" collapse. Thomas's spherically symmetric code allowed for a multi-phase fluid but no collisionless dark matter, while the code we develop here evolves a single fluid component and a single collisionless component. Shapiro & Struck-Marcell included a collisionless component, and they examined collapses with planar rather than spherical symmetry. Our treatment of radiative cooling is somewhat different from that adopted by these authors; in particular, we can include the influence of a photo-ionizing background on the abundances of ionic species. However, the largest differences are not in the codes but in the choice of problem, and the consequent choice of initial conditions.

The geometry in our calculations is idealized, and one must therefore take care to keep their limitations in mind. Nonetheless, they can provide a valuable complement to their more elaborate, 3-dimensional cousins because of their high resolution, their speed, and their relative simplicity. Three-dimensional hydrodynamic simulations of galaxy formation suffer from limited spatial resolution and mass resolution, making it difficult to separate genuine physical results from numerical artifacts. One-dimensional collapse calculations can achieve much higher resolution, computing gas dynamic processes with much higher accuracy. Because they run fast, it is possible to undertake a much more comprehensive exploration of the parameter space, varying both the values of cosmological parameters and the assumptions about the gas microphysics. The results are much easier to visualize and interpret than those of three-dimensional simulations. Thus, these simplified, high-resolution calculations can provide a useful numerical check on three-dimensional simulations and, equally important, provide physical insight into their results. They can also check and improve upon the simpler analytic models that serve as inputs to Press-Schechter type calculations.

This paper serves two purposes. First, it describes the code itself, and tests its ability to reproduce known analytic results such as the self-similar, spherical infall solution of Bertschinger (1985). Second, it applies the code to one of the basic questions of galaxy formation: what causes the abrupt cutoff at the upper end of the galaxy luminosity function? It has long been recognized that gravitational effects alone cannot explain this cutoff, because the largest virialized objects – rich galaxy clusters – have much higher masses than the largest galaxies (White & Rees 1978). The "lore", deriving largely from Binney (1977), Silk (1977), and especially RO, is that the upper cutoff is determined mainly by atomic physics, specifically by the requirement that the gas within a density perturbation be able to cool and collapse within a Hubble time. RO even include a "numerological digression" in which they relate the characteristic masses and sizes of galaxies directly to fundamental gravitational and atomic constants, independent of cosmological parameters. We will examine the underpinnings of this argument by studying the dynamics of gas in collapsing systems of various masses, focusing especially on the cooling in high-mass perturbations. In a later paper, we will examine the suggestion of Efstathiou (1992) that photo-ionization by the UV background may strongly affect the formation of *low-mass* galaxies.

This paper is organized as follows. In §2 we present our general numerical model, focusing on the treatment of hydrodynamics and cooling in the gas component. In §3, we show the results of several test calculations. In §4, we present results for collapses of spherical density pertubations, with and without a collisionless component. In §5, we discuss the implications of these results for the galaxy luminosity function, comparing our analysis to those of RO and White & Frenk (1991).

## 2. Numerical Model

The simplest model for galaxy formation consists in the evolution of a uniform, pressureless, spherical density enhancement in a Friedmann Universe. The expansion of such a region lags behind the Hubble flow, until it stops at a turnaround time $t_{ta}$ and radius $r_{ta}$, and recollapses.

It then undergoes violent relaxation and virializes after another $\sim 1 - 2t_{ta}$. The value of $t_{ta}$ depends on the amplitude of the density contrast $\delta\rho/\rho$ at the recombination epoch. The large-scale distribution of galaxies is consistent with smooth hierarchical clustering resulting from a purely gravitational process, with no preferred length-scale. However, dissipation must have played a role in the formation of galaxies themselves. Within the spherical model, one expects gas to shock and heat as it collapses, and pressure forces and radiative cooling will strongly affect the post-shock evolution.

To better understand these important effects, we have developed a simple but highly accurate *one-dimensional* numerical code to model the collapse of individual spherical perturbations. The code treats a mixture of gas particles, evolved through Lagrangian hydrodynamics, and collisionless particles (cold dark matter), each represented by concentric shells. The Lagrangian description is preferable to an Eulerian description because it follows the mass and fluid elements themselves, thereby maintaining the mass resolution throughout the calculation. As we will show in §4, the density and temperature are extremely non-uniform during the collapse, leading to widely varying timescales, both in space and in time.

### 2.1. Equations

Since we study isolated, collapsing, density perturbations, we will work with physical coordinates (rather than coordinates that are comoving with the expanding universe).

The gaseous component is described by the fluid equations for a perfect gas. These are the continuity equation,

$$\frac{d\rho_g}{dt} + \rho_g \nabla \cdot \mathbf{v_g} = 0, \tag{1}$$

the momentum equation,

$$\frac{d\mathbf{v_g}}{dt} = -\frac{\nabla p}{\rho_g} + \mathbf{g}, \tag{2}$$

the energy equation,

$$\frac{du}{dt} = \frac{p}{\rho_g^2}\frac{d\rho_g}{dt} + \frac{\Gamma - \Lambda}{\rho_g}, \tag{3}$$

and the equation of state,

$$p = (\gamma - 1)\rho_g u. \tag{4}$$

In these equations, $\rho_g$, $\mathbf{v}_g$, $p$, and $u$ are the baryonic mass density, velocity, pressure, and internal energy per unit mass, $\Gamma$ is the external heating rate (e.g. from photo-ionization), $\Lambda$ is the radiative cooling rate, $\gamma$ is the adiabatic index, and $\mathbf{g} = -\nabla\Phi$, where $\Phi$ is the total (gas and dark matter) gravitational potential.

In spherical symmetry, equations (1) and (2) can be rewritten in the Lagrangian form as

$$dm_g = 4\pi r_g^2 \rho_g dr_g, \tag{5}$$

which replaces the continuity equation, and

$$\frac{dv_g}{dt} = -4\pi r_g^2 \frac{dp}{dm_g} - \frac{M(r_g)}{r_g^2}, \qquad (6)$$

where $r_g$ is the radius of the gas shell and $M(r)$ is the total (baryonic and dark matter) mass inside radius $r$. Here and throughout this paper we have set $G = 1$.

The collisionless component is simply described by the equation of motion

$$\frac{d\mathbf{v}_d}{dt} = \mathbf{g}, \qquad (7)$$

where $\mathbf{v}_d$ is the dark matter velocity. In spherical symmetry, this can be rewritten as

$$\frac{dv_d}{dt} = -\frac{M(r_d)}{r_d^2}, \qquad (8)$$

where $r_d$ is the radius of the collisionless mass shell.

## 2.2. Radiative cooling

We compute radiative cooling for a gas of primordial composition, 76% hydrogen and 24% helium by mass. The full set of equations that we use to obtain abundances and cooling rates is listed in Katz, Weinberg & Hernquist (in preparation), so here we restrict ourselves to a brief summary of the physics. Our original source for most of these formulae is Black (1981), and we adopt the high-temperature corrections of Cen (1992).

We compute the abundances of ionic species as a function of density and temperature by assuming that the gas is in ionization equilibrium with a spatially uniform background of UV radiation. In other words, we choose the abundances so that the rate at which each species is depopulated by photo-ionization, collisional ionization, or recombination to a less ionized state is equal to the rate at which it is populated by recombination from a more ionized state or by photo-ionization or collisional ionization of a less ionized state. The intensity and spectrum of the UV background are specified as a function of time by the user, based on theoretical models or observational constraints. Given the ionic abundances and the density, we compute the cooling rates due to collisional excitation, collisional ionization, recombination, and Bremsstrahlung, and we compute the heating rate from photo-ionization. For the physical problems that we study here, the timescales for reaching ionization equilibrium are much shorter than the other timescales of interest, so our equilibrium assumption should be an excellent approximation.

For the simulations in §§4.2 and 4.3 of this paper, we set the UV radiation background to zero. In this case the relative abundances are determined by collisional equilibrium alone, and they depend only on temperature. Since all of the cooling processes that we consider involve two-body interactions, we can describe the cooling rate $\Lambda$ by a single function of temperature, up to a factor of $\rho_g^2$. The thick solid line in Figure 1 shows the total cooling rate $\Lambda/n_H^2$, where $n_H = 0.76\rho_g/m_p$ is the number density of hydrogen nuclei. Short-dashed lines show the contributions from collisional excitation of H and He$^+$, which dominate cooling in the range $10^4$K $< T < 10^{5.5}$K. Dotted lines show the contributions from collisional ionization. Long-dashed lines show the contribution from recombination. The thin solid line shows the Bremmstrahlung contribution, which dominates at $T > 10^{5.5}$K. Below $10^4$K, all of the gas is neutral, and there are no collisions energetic enough to cause electronic excitations, so none of the processes that we consider can produce any significant cooling in this regime. At very high densities the formation of hydrogen molecules can cool primordial gas to the temperatures required for fragmentation into stars. We will not attempt to resolve stellar mass scales in our cosmological studies, so we will just consider gas that cools to $10^4$K to be "cold" and leave it at that.

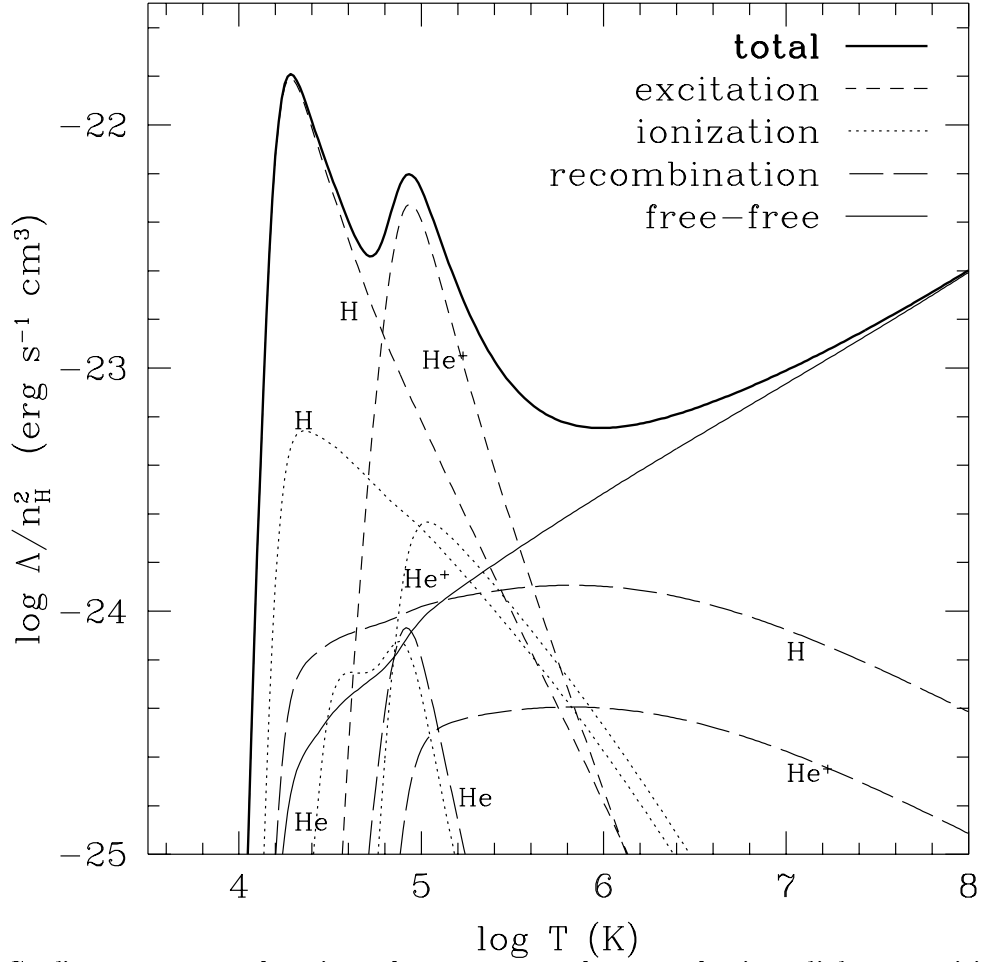

Fig. 1.— Cooling rates as a function of temperature for gas of primordial composition ($f_H = n_H m_p/\rho_g = 0.76$, $f_{He} = 0.24$). The thick solid line is the total cooling rate. The dotted lines are the cooling rates from collisional ionization of H, He, and $He^+$; the long-dashed lines are the cooling rates from recombination to H, He, and $He^+$; the short-dashed lines are the cooling rates from collisional excitation of H, He, and $He^+$; the thin solid line is the cooling rate from Bremsstrahlung. Abundances of ionic species are computed assuming collisional equilibrium.

The principal scientific concern of this paper is the cutoff at the high-luminosity end of the galaxy luminosity function. We are therefore interested primarily in the behavior of high-mass perturbations. At the virial temperatures associated with these perturbations, typically $10^6$K or greater, the gas is fully ionized by collisional processes, so our neglect of the photo-ionizing background should make no difference to our conclusions. Photo-ionization can affect the behavior of lower mass perturbations because it eliminates the neutral hydrogen and singly ionized helium that dominate cooling at low temperatures, and because the residual energy of the photo-electrons heats low-density gas to $T \sim 10^4$K. Our next paper will focus on the influence of photo-ionization on the collapse and cooling of low-mass perturbations.

Compton scattering of microwave background photons by electrons can be an important source of cooling at redshifts $z \gtrsim 10$. However, in the collapse calculations in this paper the gas remains neutral until much lower redshifts, when it collapses and shocks, so we can safely ignore Compton cooling. It would be straightforward to add this effect to our code, and we would need to do so in order to study collapses at higher redshifts.

### 2.3. Numerical scheme

We use the standard, second-order accurate, Lagrangian finite-difference scheme (Bowers & Wilson 1991). In this scheme, the velocity is zone-edge-centered, while the pressure and internal energy are zone-centered. To obtain time-centering, the velocities are evaluated at half-timesteps. We give the hydrodynamical finite-difference equations in the order in which they must be evaluated. In the following, the subscripts denote the position of the shell, and the superscripts denote the time. We first present the equations for the gas component, and for clarity of presentation we drop the subscripts $g$. Note that in the following equations, $\rho_i^n$ is the *gas* density, and $m_i^n$ is the *total* mass interior to the shell at position $i$ at time $n$, including both gas and dark matter. First we must advance the velocities to $t^{n+1/2}$ according to

$$v_i^{n+1/2} = v_i^{n-1/2} - \left[ 4\pi (r_i^n)^2 \frac{p_{i+1/2}^n - p_{i-1/2}^n}{dm_i} + \frac{m_i^n}{(r_i^n)^2} \right] dt^n. \tag{9}$$

Then we can advance the positions and evaluate the densities,

$$r_i^{n+1} = r_i^n + v_i^{n+1/2} dt^{n+1/2}, \tag{10}$$

and

$$\rho_{i+1/2}^{n+1} = \frac{dm_{i+1/2}}{\frac{4}{3}\pi \left[ (r_{i+1}^{n+1})^3 - (r_i^{n+1})^3 \right]}. \tag{11}$$

In these equations,

$$dt^n = \frac{1}{2}(dt^{n-1/2} + dt^{n+1/2}), \tag{12}$$

and

$$dm_i = \frac{1}{2}(dm_{i-1/2} + dm_{i+1/2}). \tag{13}$$

Equations (9)-(11) are second-order accurate in space and time. We can now advance the energy equation:

$$u_{i+1/2}^{n+1} = u_{i+1/2}^n - p_{i+1/2}^n \left( \frac{1}{\rho_{i+1/2}^{n+1}} - \frac{1}{\rho_{i+1/2}^n} \right) + \frac{(\Gamma - \Lambda)_{i+1/2}^n}{\rho_{g,i+1/2}^{n+1}} dt^{n+1/2}. \tag{14}$$

This last equation is only first-order accurate in time. To make it second-order accurate, we would have to replace $p_{i+1/2}^n$ and $(\Gamma - \Lambda)_{i+1/2}^n$ by $p_{i+1/2}^{n+1/2}$ and $(\Gamma - \Lambda)_{i+1/2}^{n+1/2}$. This would require two evaluations of the cooling function per timestep, and since a lot of the computational time is in practice spent evaluating the cooling functions, this would be quite expensive. However, as we will show in §3, the energy is extremely well conserved in our present scheme, with equation (14), and there is therefore no reason to require more accuracy in the energy equation.

Shocks are treated with the usual artificial viscosity technique (Richtmyer & Morton 1967). The pressure in the momentum and energy equations is replaced by $P = p + q$, where

$$q_{i+1/2}^{n+1} = -c_q \frac{2}{(1/\rho_{i+1/2}^{n+1}) + (1/\rho_{i+1/2}^n)} |v_{i+1}^{n+1/2} - v_i^{n+1/2}| (v_{i+1}^{n+1/2} - v_i^{n+1/2}) \tag{15}$$

if $v_{i+1}^{n+1/2} - v_i^{n+1/2} < 0$ and $q = 0$ otherwise. We use $c_q = 4$, which spreads shock fronts over 4-5 shells.

The second-order accurate, finite-difference equations for the collisionless equation of motion are much simpler than those for the gas,

$$v_{d,i}^{n+1/2} = v_{d,i}^{n-1/2} - \frac{m_i^n}{(r_{d,i}^n)^2}, \tag{16}$$

and

$$r_{d,i}^{n+1} = r_{d,i}^n + v_{d,i}^{n+1/2} dt^{n+1/2}. \tag{17}$$

In the absence of shell-crossing, the mass $m_i$ inside a given shell would be constant in time. However, the collisionless shells are allowed to cross each other and to cross gas shells, so the masses $m_i^n$ are functions of time and must be computed at each time step.

### 2.4. Timesteps and central boundary conditions

For each shell in the calculation there are three potentially important timescales, namely the dynamical, Courant, and cooling timescales. In addition to respecting these timescales, we must ensure that the fluid shells do not cross. We therefore take a timestep

$$dt = \min \{dt_{dyn}, dt_{Cour}, dt_{cool}, dt_{vel}\}, \qquad (18)$$

where

$$dt_{dyn} = \min_i \left\{ c_d \sqrt{\frac{\pi^2 r_i^3}{4 m_i}} \right\}, \qquad (19)$$

$$dt_{Cour} = \min_i \left\{ c_C \left| \frac{r_i - r_{i-1}}{\sqrt{\gamma(\gamma-1)u_i}} \right| \right\}, \qquad (20)$$

$$dt_{cool} = \min_i \left\{ c_c \left| \frac{u_i \rho_i}{(\Gamma - \Lambda)_i} \right| \right\}, \qquad (21)$$

and

$$dt_{vel} = \min_i \left\{ c_v \left| \frac{r_i - r_{i-1}}{v_i - v_{i-1}} \right| \right\}, \qquad (22)$$

where $c_d$, $c_C$, $c_c$, and $c_v$ are safety constants. We use $c_d = 0.01$, $c_C = 0.2$, $c_c = 0.1$, and $c_v = 0.05$. The timesteps vary widely both in space and in time during a calculation. It is therefore essential to compute appropriate timesteps for all shells in the calculation at a given time, and to use the smallest of these to advance the system. We could increase computational efficiency at the price of additional complication by allowing different shells to have different timesteps, but thus far we have not found it necessary to adopt this procedure.

A Lagrangian code achieves very high spatial resolution near the center, and it can thereby demand extremely short timesteps. For collisionless particles the timestep is determined by $dt_{dyn} \propto \sqrt{r^3/m(r)}$. The first shell, with mass $dm$, has $dt_{dyn} \propto \sqrt{r_1^3/dm} \propto r_1^{3/2}$, so the timestep goes to zero as the shell approaches $r = 0$. Since spherical symmetry is an idealization in any case, there is no need to bring the calculation to a grinding halt in order to integrate the very central region at high accuracy. We therefore solve the timestep problem by treating the center as a hard reflecting sphere of radius $r_c$ (Spitzer & Hart 1971; Gott 1975), which prevents timesteps from becoming arbitrarily small. Energy conservation degrades as $r_c$ is decreased (because shells rebound at higher velocities), but it is not difficult to find a value of $r_c$ that (a) yields excellent energy conservation and (b) is much smaller than other characteristic radii in the problem, so that the departure from an idealized spherical collapse is minimal.

A similar problem arises for the gaseous component when the gas elements cool very rapidly. In regions where the cooling time is much smaller than the dynamical time, the fluid shells collapse at the free-fall rate, the central density increases very rapidly, and the cooling and dynamical timesteps become exceedingly small. We solve the problem by "freezing" shells that have $t_{cool} \ll t_{dyn}$. More specifically, when a shell has $t_{cool} < c_f t_{dyn}$, we move it to a radius $r = r_{min}$, assign it zero velocity and a temperature of $10^4$ K, and subsequently ignore it, except in the calculation of the gravitational force. We require $(r - r_{min}) < 0.1 r_{min}$, so that no shell moves a large fraction of its radius in the freezing process; if the condition $t_{cool} < c_f t_{dyn}$ is reached at a large radius, we continue to evolve the shell until it reaches $r < 1.1 r_{min}$. Shells in this regime collapse nearly isothermally at the free-fall rate. We have run tests with $c_f = 10^{-2}$ and $c_f = 10^{-3}$; these values yield virtually identical results. We adopt $r_{min} = r_c$, the radius of the reflecting sphere used for the collisionless component. We first tried moving frozen shells to the origin, but this practice leads to unstable numerical results because the shells have a wide range of values of $t_{dyn}$ at the time they are frozen. Adopting $r_{min} = r_c$ suppresses this instability, though in one physical regime (where a large fraction but not 100% of a perturbation collapses), the amount of cooled mass can vary by $\sim 10\%$ from one run to another, depending on the number of shells used, because of residual sensitivity to the central boundary condition. Such variations are small compared to the approximation made in treating galaxy formation as a spherically symmetric process.

## 3. Test Calculations

We have performed a variety of tests of the code on problems with known analytic solutions. We describe results from several of these tests in this section.

### 3.1. Polytropes

In the absence of radiative cooling, the total energy of the system must be conserved, and in the absence of shocks, the entropy must be conserved. To check that the code satisfies these basic requirements, we simulate an equilibrium polytrope of index $n_p = 1.5$ ($\gamma = 5/3$). We use $N_g = 500$ shells, equally spaced in radius $r$, and units such that $G = M = R = 1$. We relax the polytrope to an equilibrium by adding a dissipative term in the momentum equation of the form $-v_g/t_{relax}$. We then remove this term and allow the polytrope to evolve dynamically. The result is an exceedingly accurate equilibrium: positions of the fluid shells fluctuate with amplitudes $\Delta r/r$ smaller than $10^{-7}$, the kinetic energy remains smaller than $10^{-14}$, and the potential energy, thermal energy, and entropy change by less than $10^{-8}$ over $\sim 100$ dynamical times.

For a more rigorous trial, we set up initial conditions corresponding to the first two normal pulsation modes of the polytrope. To obtain the appropriate initial configuration, we simultaneously integrate the Lane-Emden equation and the eigenvalue equation governing the radial oscillations of a spherical star (Shapiro & Teukolsky 1983). We choose an initial mode amplitude equal to 1% of the star radius. The eigenfunctions are shown in Figure 2a. The eigenfrequencies corresponding to these eigenfunctions are $\omega_1 = 1.645$, and $\omega_2 = 3.547$. We let the pulsating polytrope evolve dynamically, using $N_g = 500$ shells in each case. The total energy and entropy vary by less than $5 \times 10^{-5}$ over about 20 periods of oscillation. The oscillations of the gravitational, thermal, and kinetic energies are shown in Figure 2b. The oscillation periods are $T_1 = 3.82$ and $T_2 = 1.77$ for the first and second modes, respectively. These are identical to the periods $2\pi/\omega$ obtained from the eigenfrequencies of these modes.

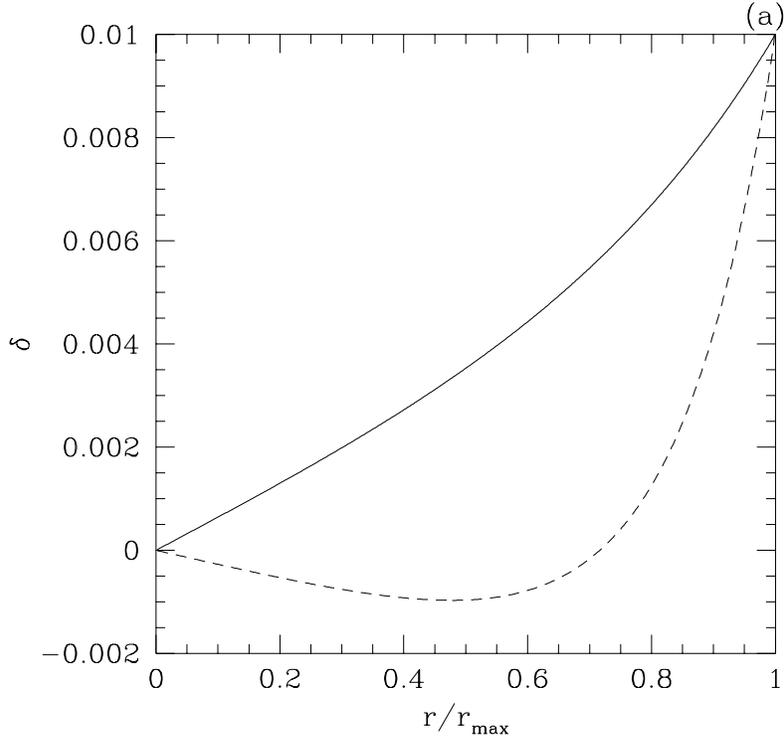

Fig. 2.— (a) Eigenfunctions for the first two pulsating modes of a polytrope with adiabatic exponent $\gamma = 5/3$. To set up a pulsating polytrope we perturb the shell radii of the equilibrium model according to $r \to r + \delta$. (b) Fluctuations in the total energy, the gravitational potential energy, the internal energy, and the kinetic energy, from a simulation with $N_g = 500$ shells, during 20 periods of the second normal mode. The bottom panel shows entropy production during the oscillations. (c) Fluid shell trajectories and entropy production when the second normal mode has an amplitude of 15% of the star radius.

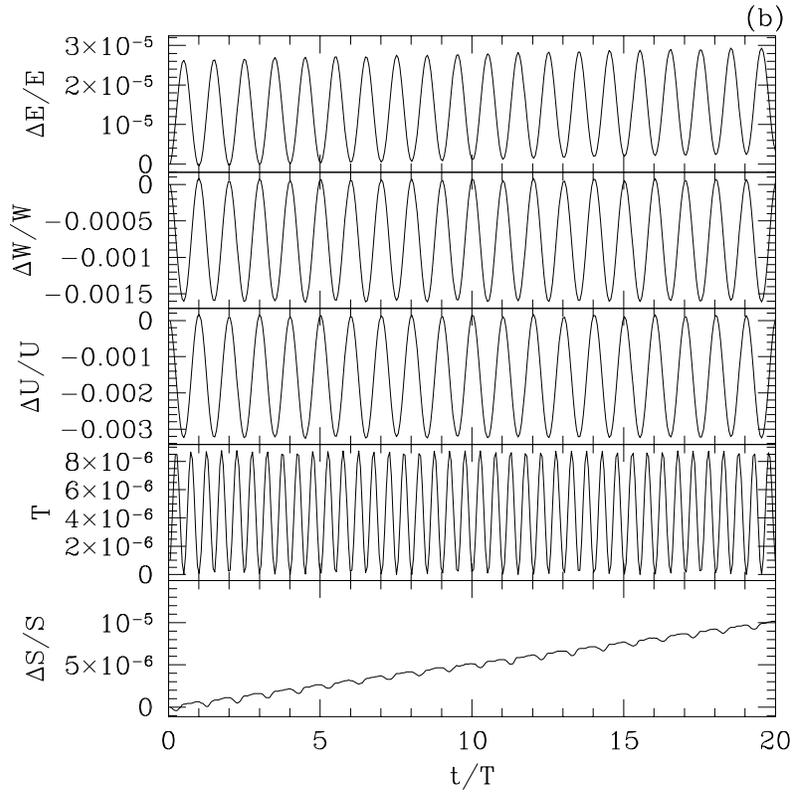

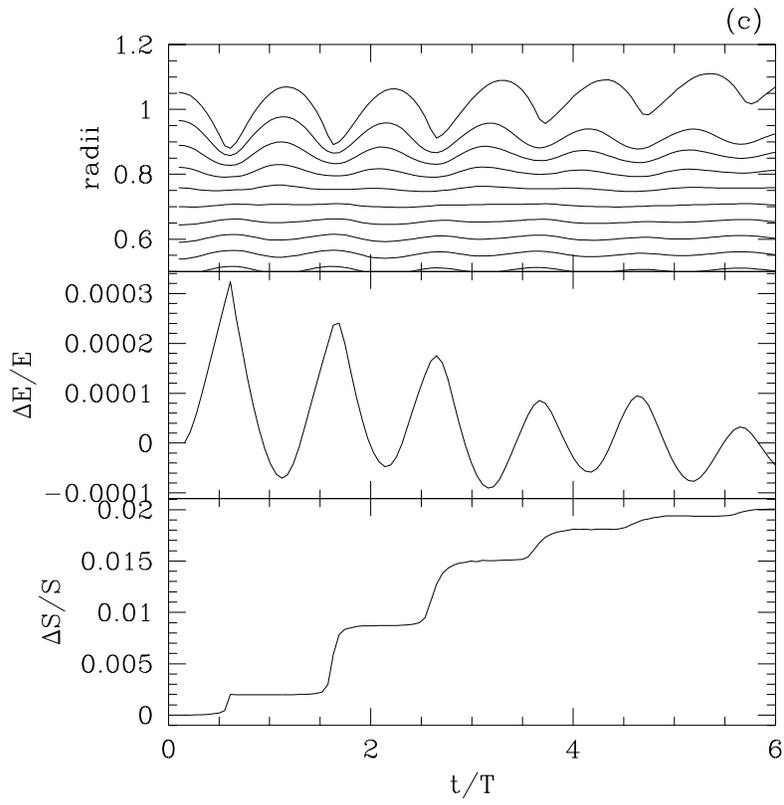

From Figure 2b we see that the entropy grows slowly during evolution. This entropy increase arises because of the finite amplitude of the modes. If we increase the mode amplitude further, the oscillations produce shocks, which generate entropy. Figure 2c shows this entropy generation when the second normal mode has an amplitude of 15% of the star radius. The shock appears in the outer shells, where the amplitude of the perturbation is the largest.

### 3.2. Self-similar hydrodynamic collapse

To test the code in a context closer to its cosmological purpose, we check that it reproduces the similarity solution for shocked accretion of a collisional, non-radiative gas, as described by Bertschinger (1985). This solution describes the asymptotic (late-time) behavior of a spherically symmetric collapse about an initial seed perturbation in an Einstein-de Sitter ($\Omega = 1$) universe. In the absence of cooling, neither the gas physics nor the cosmology defines a preferred scale, so once the collapsed mass exceeds that in the initial perturbation, the system "forgets" the details of its initial state, and its evolution becomes self-similar in time. We evolve a perturbation with a Gaussian initial density profile,

$$\delta_i(r) = \delta_i(r = 0)e^{-r^2/R_i^2}, \tag{23}$$

and an unperturbed, Hubble-flow velocity profile,

$$v_i(r) = H_i r, \qquad H_i = 2/(3t_i). \tag{24}$$

Here $t_i$ is the initial time, $H_i$ is the Hubble constant, and

$$\delta_i(r) = [\rho_i(r) - \overline{\rho}_i]/\overline{\rho}_i \tag{25}$$

is the initial density contrast, with

$$\overline{\rho}_i = \rho_{Hi} \equiv (6\pi t_i^2)^{-1}, \tag{26}$$

the critical density at time $t_i$. In the linear regime, the density contrast grows as

$$\overline{\delta} = \overline{\delta}_i \left( \frac{3}{5}\tau^{2/3} + \frac{2}{5}\tau^{-1} \right), \tag{27}$$

where

$$\overline{\delta}(r) \equiv \frac{\int_0^r \delta(r')4\pi r'^2 dr'}{\frac{4}{3}\pi r^3} \tag{28}$$

is the averaged overdensity interior to the shell at position $r$, and

$$\tau = t/t_i \tag{29}$$

is the time in units of the initial time $t_i$. The second term in equation (27), and the 3/5 factor multiplying the first term, appear because our Hubble-flow initial conditions contain a mixture of growing and decaying modes.

As the density contrast inside a shell grows, the extra gravitational deceleration drags it further behind the Hubble flow, until it finally turns around and recollapses. Roughly halfway back to the center, it hits a shock, which sharply raises its density, temperature, and entropy, and brings it nearly to rest. Thereafter, the shell falls *very* slowly towards $r = 0$. Figure 3 shows the trajectory of a typical shell in our simulation; the radius and time are scaled to the shell's turnaround radius $r'_{ta}$ and turnaround time $t_{ta}$. With this scaling, the trajectories of all shells that lie well outside the initial perturbation are identical, and the trajectory shown in Figure 3 is indistinguishable from that in figure 4 of Bertschinger (1985).

Figure 4 shows the velocity, density, pressure, and mass profiles from a simulation with $\delta_i(r = 0) = 0.2$, $R_i = 1$, a $\gamma = 5/3$ equation of state, and $N_g = 1000$ shells. Solid lines show results at $\tau = 1000, 2000, 3000, 4000$, and 5000, from bottom to top. The dashed line, often obscured by the solid lines, shows Bertschinger's (1985) similarity solution. The dimensionless radius, velocity, mass, density, and pressure are defined by

$$\lambda = r/r_{ta}, \tag{30}$$

$$V(\lambda) = v(r,t)(r_{ta}/t)^{-1}, \tag{31}$$

$$M(\lambda) = m(r,t)\left(\frac{4}{3}\pi\rho_H r_{ta}^3\right)^{-1}, \tag{32}$$

$$D(\lambda) = \rho(r,t)/\rho_H, \tag{33}$$

and

$$P(\lambda) = p(r,t)(t/r_{ta})^2 \rho_H^{-1}. \tag{34}$$

Here $r_{ta}$ is the radius of the shell that is currently turning around (and is thus distinct from the radius $r'_{ta}$ used in Figure 3). We see in Figure 4 that the profiles for the dimensionless fluid parameters tend asymptotically towards the similarity solution, demonstrating the ability of our code to reproduce known analytic results for this problem, one that is directly relevant to the cosmological applications that we will consider. The total energy is conserved to better than 0.2% over the entire run.

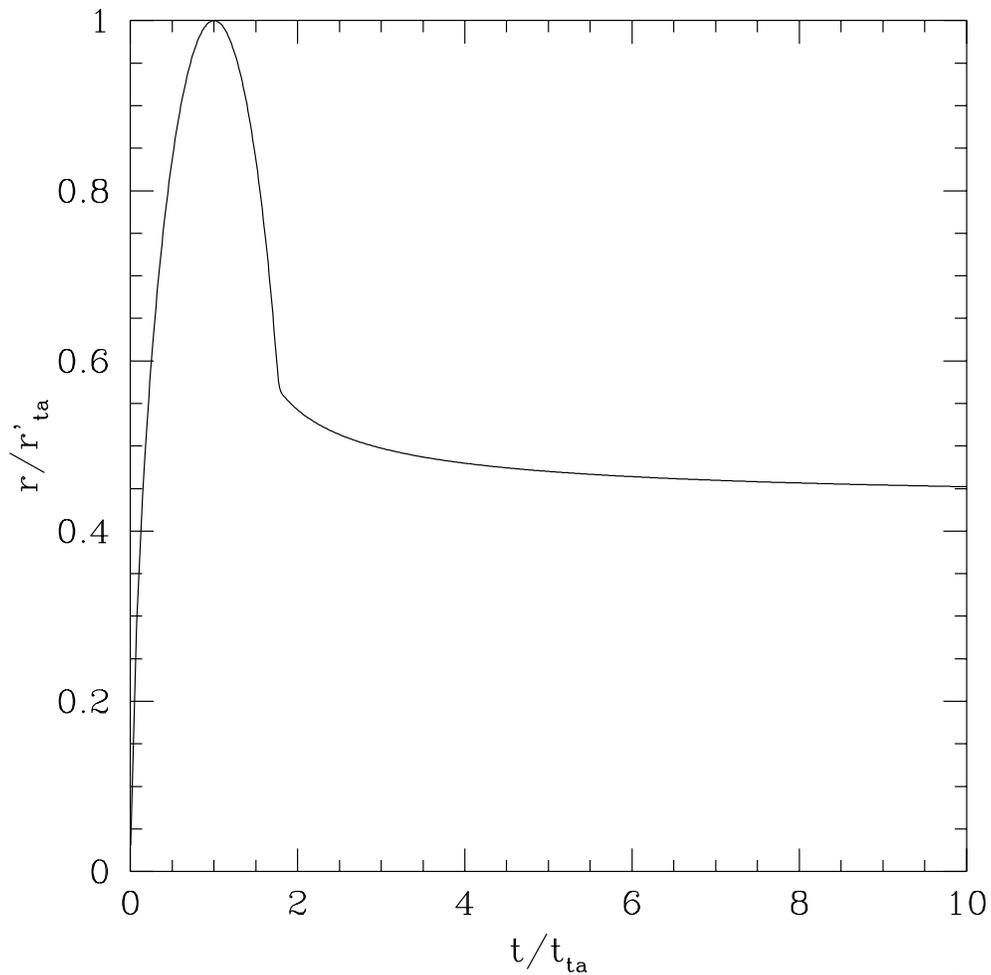

Fig. 3.— Trajectory of a fluid shell in the self-similar, shocked accretion of a $\gamma = 5/3$ collisional gas. The radius and time are scaled to the shell's turnaround radius $r'_{ta}$ and turnaround time $t_{ta}$. With this scaling, the trajectories are identical for all shells.

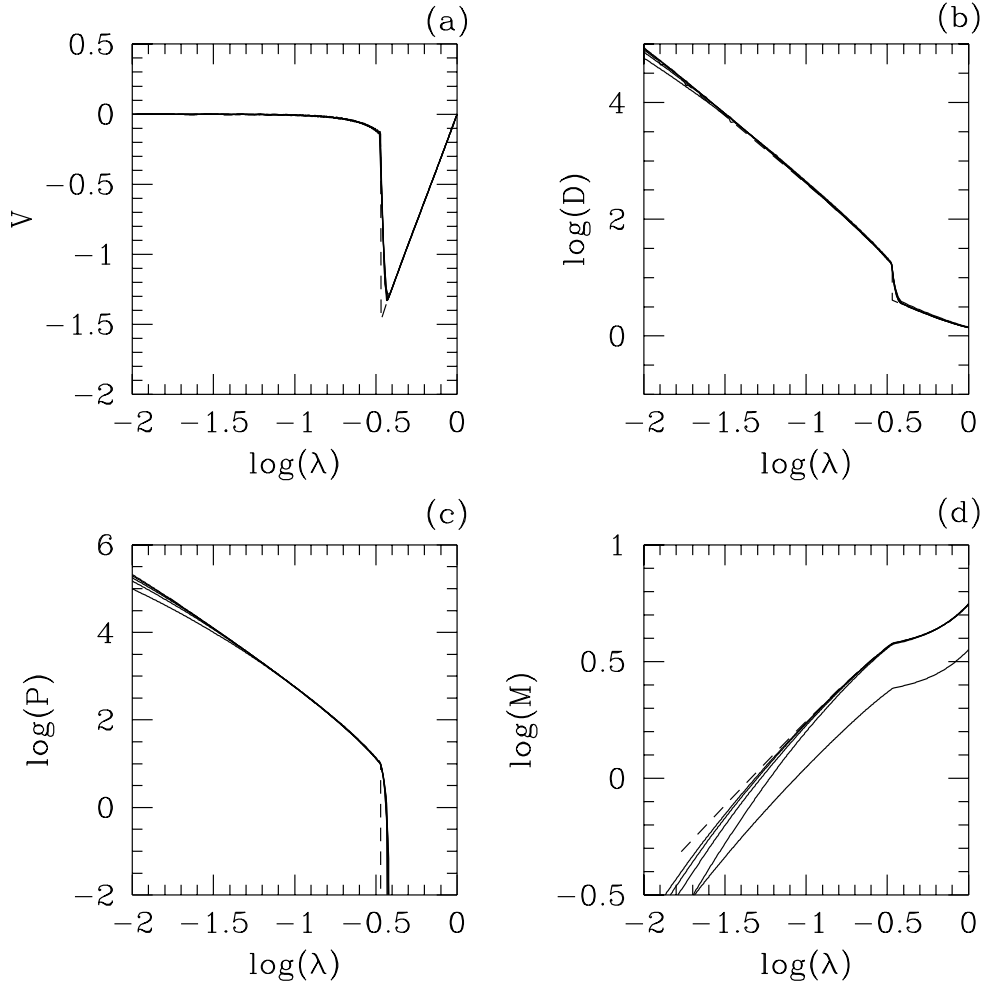

Fig. 4.— Convergence of the numerical results toward the similarity solution for shocked accretion of a $\gamma = 5/3$ collisional gas. The non-dimensional velocity $V$, density $D$, pressure $P$, and mass $M$ are shown as a function of the scaled radius $\lambda$. The units are given by eqns. (30)-(34). Profiles from a simulation with $N_g = 1000$ shells are plotted at $\tau = 1000, 2000, 3000, 4000,$ and $5000$, from bottom to top, where $\tau$ is defined in eqn. (29). The dashed line shows the similarity solution from Bertschinger (1985).

### 3.3. Pressureless collapse onto a black hole

In the case of cold accretion onto a black hole, there is no shell-crossing, $M(r)$ is constant in time, and the equation of motion $\ddot{r} = -M(r)/r^2$ can be integrated analytically for each shell. Integrating once we get $v^2/2 = M/r - C$, where $C = M/r_i - v_i^2/2$, and $r_i$ and $v_i$ are the position and velocity of the shell at the initial time $t_i$. This equation can be integrated once more to give the radii as function of time, yielding the well-known cycloid solution, usually written in parametric form (see, e.g., Padmanabhan 1993, §8). The mass profile and the velocity profile are therefore known analytically.

For this test, we start with a top-hat initial density profile,

$$\rho_i = \rho_{Hi} \begin{cases} 1 + \delta_i & \text{if } r < R_i \\ 1 & \text{if } r > R_i, \end{cases} \quad (35)$$

with $\delta_i = 0.3$ and $R_i = 1$. We evolve the collisionless system forward in time and compare the numerical results for the mass and velocity profiles to the exact analytical values. In Figure 5a and 5b, solid lines show the numerical velocity and mass profiles, obtained for $N_d = 10,000$ shells. Points show the exact results at selected values of $\lambda$. Dashed lines represent the unperturbed Hubble flow. Numerical and analytic results agree to better than 0.5%, as shown in Figures 5c and 5d.

At the time shown in Figure 5, the collapsed mass significantly exceeds the mass in the initial top-hat perturbation, so the system has reached the regime of self-similar evolution. Our Figures 5a and 5b are, therefore, identical to the similarity solution plotted in figure 1 of Bertschinger (1985).

### 3.4. Coupled hydrodynamic and collisionless systems

One important new numerical effect enters when we model mixed collapses: as collisionless shells cross fluid shells they introduce discrete fluctuations in the gravitational forces acting on the fluid, and this spurious agitation leads to low-level shocks and associated entropy production. The discreteness effect depends primarily on the mass ratio between collisionless and fluid shells. From a variety of tests, we find that this effect becomes negligible when the mass ratio is unity or smaller. In general, therefore, we require $N_d/N_g \geq \Omega_d/\Omega_g$.

### 4. Collapses with radiative cooling

We now turn to the main scientific issue of this paper: the relation between the physics of gas cooling and the rather sharp upper cutoff in the distribution of galaxy luminosities. We address this point by modeling spherical collapses with radiative cooling appropriate to a gas of primordial composition in collisional equilibrium (see §2.2 and Figure 1).

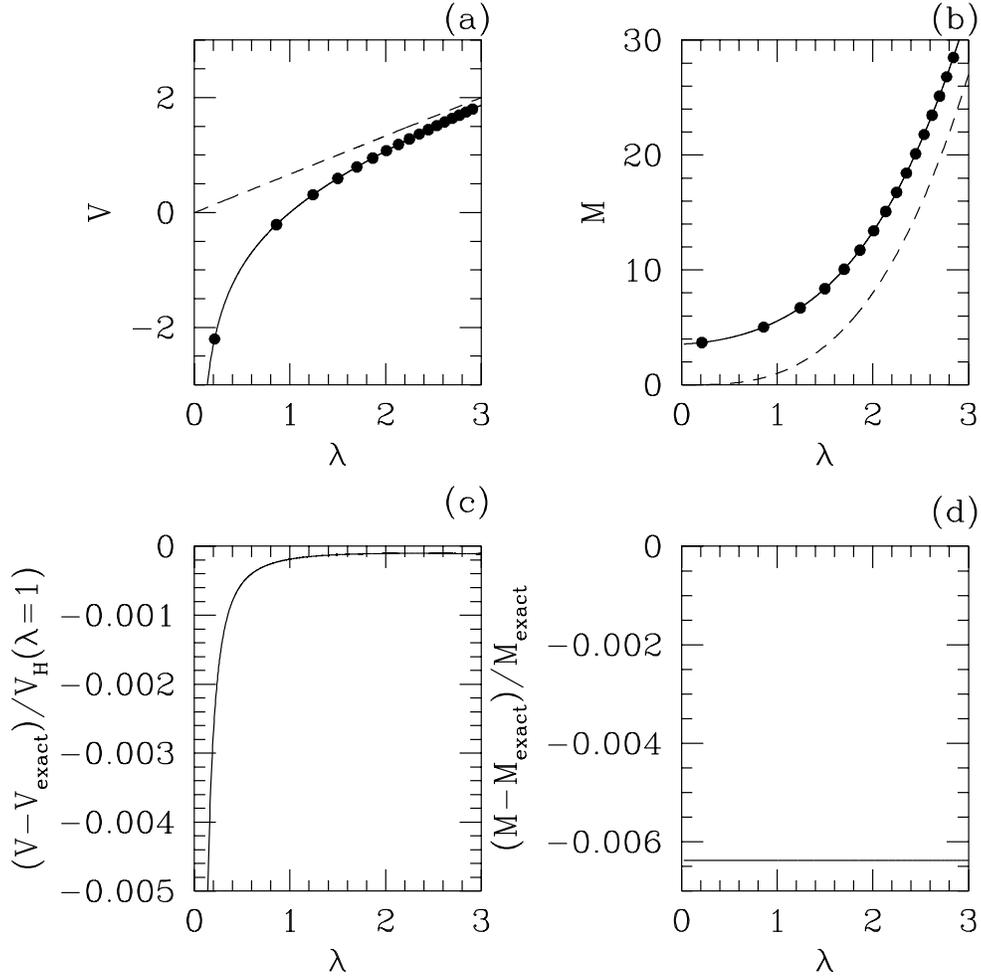

Fig. 5.— Numerical results (solid lines) for pressureless collapse onto a black hole, from a simulation with $N_d = 10,000$ collisionless shells. Upper panels show the non-dimensional velocity $V$ and mass $M$ as a function of the scaled radius $\lambda$. The units are given by eqs. (30)-(33). Points show exact analytic results, and dashed lines show the result for unperturbed Hubble flow. Lower panels show the error in the numerical results. The velocity error is scaled to a characteristic velocity of the problem, $v_H(\lambda = 1)$.

### 4.1. Initial conditions

As initial conditions, we adopt the average density profile around a peak in a Gaussian random density field. Equation (7.10) of Bardeen et al. (1986) gives the orientation-averaged, mean density profile around a peak of height $\nu\sigma_i$ and curvature $x$,

$$\frac{F(r)}{\sigma_i} = \frac{\nu}{(1-\beta^2)}(\psi + \frac{\nabla^2\psi}{3}) - \frac{x/\beta}{(1-\beta^2)}(\beta^2\psi + \frac{\nabla^2\psi}{3}), \tag{36}$$

where $\beta$ is a constant that depends on the slope of the power spectrum, $\psi(r) \equiv \xi(r)/\xi(0)$ is the normalized correlation function, and $\sigma_i$ is the rms density fluctuation. For a density field with a power-law power spectrum of index $n$ smoothed by convolution with a Gaussian filter of radius $R_f$, the power spectrum is $P(k) = Ak^n e^{-k^2 R_f^2}$, and the normalized correlation function, its Fourier transform, is

$$\psi(r) = \phi(\frac{3+n}{2}, \frac{3}{2}; \frac{-r^2}{4R_f^2}), \tag{37}$$

where $\phi$ is the degenerate hypergeometric function (defined in Gradshteyn & Ryzhik 1980, §9.210.1). In this paper we adopt $n = -2$, roughly the slope of the cold dark matter power spectrum on galactic scales. For this choice, $\phi$ is related to the gamma function through $\gamma(a, x) = (x^\alpha/\alpha)\phi(\alpha, 1+\alpha; -x)$. The coefficient $\beta = \sqrt{(n+3)/(n+5)} = 1/3$, and

$$<x> = \beta\nu + \frac{3(1-\beta^2) + (1.216 - 0.9\beta^4)e^{-\beta/2(\nu\beta/2)^2}}{[3(1-\beta^2) + 0.45 + (\beta\nu/2)^2]^{1/2} + \beta\nu/2}. \tag{38}$$

Figure 6 shows the normalized profile, $F(r)/\sigma_i$, and the spherically averaged interior overdensity, $\overline{F(r)}/\sigma_i$, for a $2\sigma_i$ peak. It is this profile that we adopt as the initial density distribution for our simulations. We do not expect our qualitative results to be sensitive to the details of this choice. The perturbed initial velocity profile is that implied by the growing mode solution of linear theory, $v_i = H_i r_i(1 - \overline{\delta}_i/3)$.

Given the profile shape of Figure 6, our initial conditions have two free parameters: the filter radius $R_f$ and the initial overdensity of the peak at $r = 0$, $\delta_p$. Physically, it is more convenient to describe the perturbations in terms of a mass scale and a redshift of collapse. We define the filter mass $M_f$ to be the mass contained within a sphere of radius $2R_f$, and we define the collapse redshift $z_c$ to be the redshift at which the $r = 2R_f$ shell would collapse to $r = 0$ in the absence of pressure. For our adopted profile, $\overline{\delta}(2R_f)/\delta_p = 0.6805$, so the mass $M_f$ is related to the filter radius $R_f$ by

$$\begin{align}M_f &= \tfrac{4\pi}{3}(2R_f)^3 \rho_{H,i}[1 + \overline{\delta}_i(2R_f)] \tag{39}\\ &= \tfrac{2}{9t_i^2}(2R_f)^3(1 + 0.6805\delta_p). \tag{40}\end{align}$$

Specifying the initial time and the collapse redshift $z_c$ determines the value of $\delta_p$,

$$\delta_p = \overline{\delta}_i(2R_f)/0.6805 = 1.69\left(\frac{1+z_c}{1+z_i}\right)/0.6805. \tag{41}$$

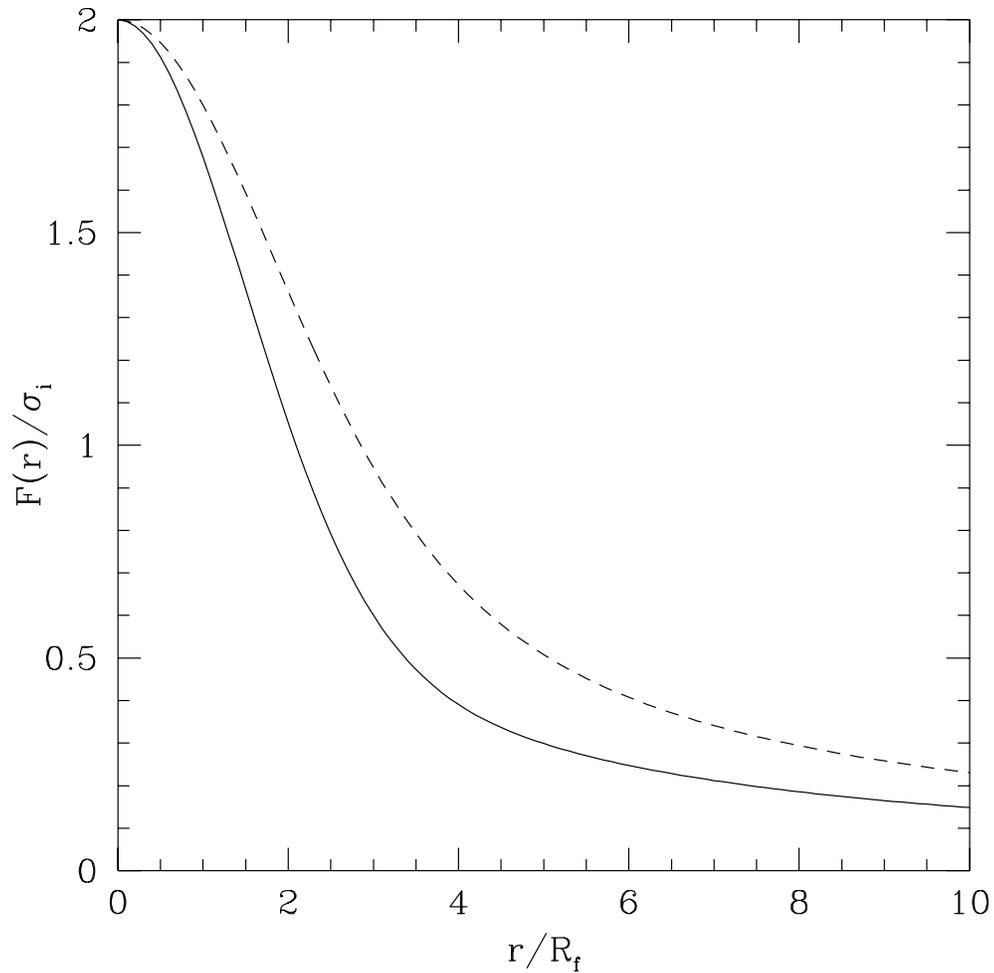

Fig. 6.— Mean profile of a $2\sigma$ peak in a Gaussian density field with power spectrum $P(k) = Ak^{-2}e^{-k^2 R_f^2}$. The solid line shows the density contrast; the dashed line shows the mean interior density contrast, as defined in eqn. (28). We adopt this profile as the initial density distribution for our collapse simulations.

Equation (41) is specific to an $\Omega = 1$ universe, but the generalization to an open universe is straightforward.

By varying the mass $M_f$ collapsing at a fixed redshift $z_c$, we effectively vary the filter radius $R_f$. By varying the redshift at which a given mass collapses, we vary the density at which the collapse occurs. In practice, we want to ensure that simulations start in the linear regime, so we fix the initial peak overdensity at $\delta_p = 0.2$. Specifying $z_c$ determines the initial redshift through equation (41), and specifying $M_f$ determines $R_f$ through equation (40). The initial redshift is $z_i \approx 36$ for $z_c = 2$ and $z_i \approx 24$ for $z_c = 1$.

### 4.2. Pure fluid collapses

In the absence of cooling, physical processes do not introduce any preferred length or time scales. In appropriate units, therefore, the results of a collapse are the same for all values of $M_f$ and $z_c$: radii scale as $r \propto M_f^{1/3}(1+z_c)^{-1}$, and times scale as $t \propto (1+z_c)^{-3/2}$. Figure 7 shows the $r$ vs. $t$ trajectories of several fluid shells in a collapse calculation without cooling, using the initial density profile described in §4.1 and $N_g = 300$ fluid shells. Shell radii are scaled to the initial filter radius $R_f$, and times are scaled to the collapse time $t_c$ of the shell with initial radius $2R_f$. In these units, the fluid trajectories are identical for any values of $M_f$ and $z_c$. Each shell expands initially, then turns around and recollapses until it is halted by a shock, usually when it has fallen back to about half of its maximum radius. The shock forms at the center when the first shells collapse, and it propagates outward in both Eulerian and Lagrangian coordinates. When a shell hits the shock, its fluid density increases (by a factor of four in the limit of a strong shock), and nearly all of its kinetic energy is converted to thermal energy. At $r_s = r_{ta}/2$, the pre-shock infall velocity is $v = (GM/r_s)^{1/2}$, making the post-shock temperature $T \approx (GM/r_s)(\mu m_p/3k)$, roughly the virial temperature implied by the interior mass $M$ and the shock radius $r_s$. After the shock, the shell is almost supported by the pressure of the hot gas beneath it, but as more gas piles on top it is compressed very slowly towards $r = 0$. The trajectories of the outer shells in Figure 7 are the same as the shell trajectory for the similarity solution (Figure 3), but their shape appears somewhat different because they are plotted in different units.

Radiative cooling introduces a new timescale into the collapse problem. Prior to shocking, pressure is unimportant, so a shell follows the same trajectory as before. However, the behavior of the shell after it hits the shock depends critically on the post-shock density and temperature, specifically on the ratio of the cooling time, $t_{cool} \sim u\rho/\Lambda$, to the dynamical time, $t_{dyn} \sim (G\bar{\rho})^{-1/2}$, where $\bar{\rho}$ is the mean mass density interior to the shell. While the local density $\rho$ and the mean interior density $\bar{\rho}$ are not identical, they are roughly proportional to each other. At fixed temperature, the cooling rate $\Lambda \propto \rho^2$ (assuming collisional equilibrium), so $t_{cool}/t_{dyn} \sim \rho^{-1/2}$. A shell that collapses earlier, when the density is higher, cools more efficiently. The post-shock temperature determines the initial location of the shell on the cooling curve (Figure 1), and because the cooling curve is a complicated function of temperature, the influence of the post-shock temperature on shell behavior is itself rather complicated. In particular, it is important to recall that the radiative cooling cuts off very sharply at $T \approx 10^4$K.

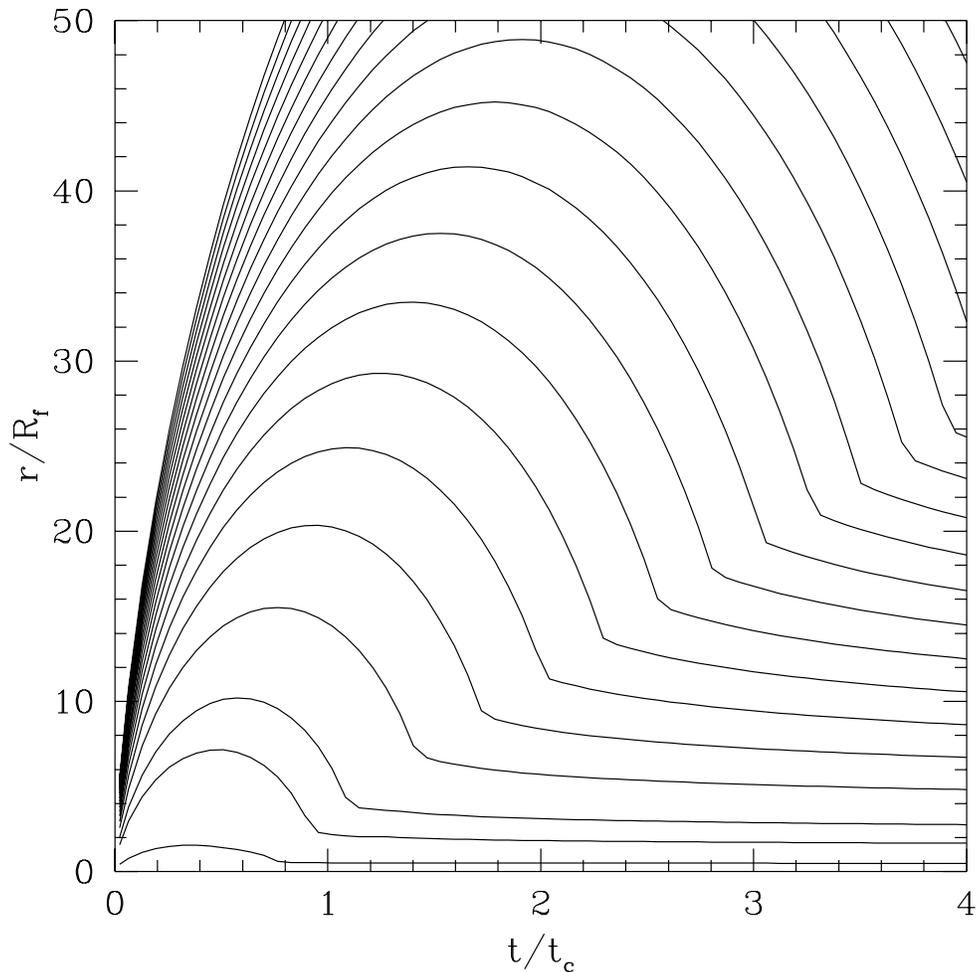

Fig. 7.— Shell trajectories for pure fluid ($\gamma = 5/3$) collapses, in the absence of cooling. The time and radii are scaled with respect to the collapse time $t_c$ and the filter radius $R_f$, defined in §4.1. With this scaling, the trajectories are independent of $M_f$ and $z_c$.

We have performed a series of collapse calculations in which we vary the value of the filter mass $M_f$ while keeping the collapse redshift fixed at $z_c = 2$. Each calculation uses $N_g = 1000$ shells. The minimum radius (see §2.4) is set to 0.99 times the initial radius of the innermost shell, making $r_c = 0.4 R_f$. We performed tests where we used $N_g = 5000$ shells and $r_c = 0.2 R_f$. The results were essentially identical.

Figures 8–10 illustrate four quite different histories of individual shells from these calculations, with Figure 8 showing the $r$ vs. $t$ trajectories, Figure 9 the trajectories in the $n - T$ plane, and Figure 10 the time evolution of the temperature and the timescale ratio $t_{cool}/t_{dyn}$. In Figures 8 and 10, $t_0$ is the age of the universe at $z = 0$. The collapse redshift $z_c = 2$ corresponds to $t = 0.192 t_0$. In Figure 9, $\overline{n}_H$ is the mean hydrogen number density inside the shell radius, $\overline{n}_H = f_H \overline{\rho}/m_p$, with $f_H = 0.76$ the mass fraction of hydrogen. The dotted lines in this Figure show contours on which the ratio $t_{cool}/t_{dyn}$ is constant, demarcating regions where cooling is rapid or slow relative to the dynamical time. The structure of these contours is closely related to the structure of the cooling curve itself. Dashed lines indicate contours in the $n - T$ plane along which the cooling time or the dynamical time is equal to the age of the universe. Diagonal solid lines are lines of constant Jeans mass, $M_J = (\pi k_B/Gm_p\mu)^{3/2} T^{3/2} \rho^{-1/2}$.

Because Figure 9 plots the quantity $\overline{n}_H$ corresponding to the mean interior overdensity, which is the relevant parameter for the dynamical time, the computation of $t_{cool}$, which depends on the *local* density, is somewhat ambiguous. In the similarity solution the ratio of local density in a recently shocked shell to the mean interior density is about a factor of four. With this result in mind, we have computed $t_{cool}(\overline{n}_H, T)$ in Figure 9 on the assumption $n_H = 4\overline{n}_H$. While this approximation is better than $n_H = \overline{n}_H$, one should still take the contours in Figure 9 as indicative rather than precise boundaries. The cooling times in Figure 10 are computed from the shell's local density, so the $t_{cool}/t_{dyn}$ ratios plotted there are more reliable.

Our four illustrative fluid shells are labeled A–D in these Figures. In Figure 8 the dot-dash line shows the trajectory of a shell with no cooling for comparison. Shell A has an interior mass of $6.3 \times 10^7 M_\odot = 2.9 M_f$. Its post-shock temperature is smaller than $10^4$K, so after the shock the shell is unable to cool, and it remains nearly pressure supported. However, the continuing infall of gas from larger radii compresses the shell, increasing its density and temperature adiabatically. The ratio of cooling time to dynamical time is large during this phase of evolution (Figure 10). Eventually, compression pushes the shell temperature above $10^4$K; the cooling time drops rapidly, and the shell collapses to $r = 0$ at the free-fall rate. This phase of the collapse is effectively isothermal, since any energy gained during compression is immediately radiated away. The points on trajectory A in Figure 9 are spaced at equal time intervals $\Delta t = 0.1 t_0$. One can see from their locations that the initial post-shock evolution is very slow but the final collapse very rapid. All points above $\log \overline{n}_H = -1$ in Figure 9 belong to trajectory B; the next point for shell A lies off the top of the plot.

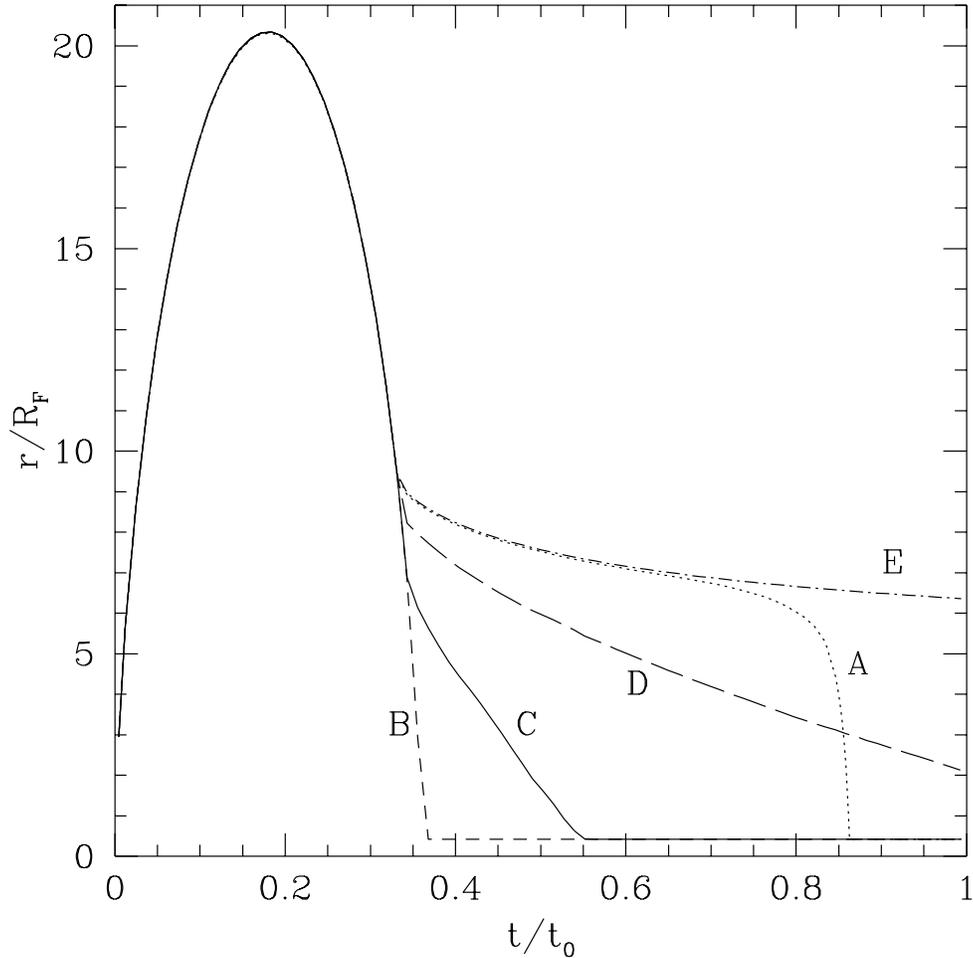

Fig. 8.— Trajectories of fluid shells in pure fluid collapses with radiative cooling, for collapse redshift $z_c = 2$ and four different values of $M_f$. Dimensionless radii $r/R_f$ are plotted as functions of the dimensionless time $t/t_0$, where $t_0$ is the age of the universe at $z = 0$. The masses interior to the shells are $6.3 \times 10^7 M_\odot = 2.9 M_f$ (shell A), $3.2 \times 10^8 M_\odot = 3.2 M_f$ (shell B), $3.2 \times 10^{14} M_\odot = 3.2 M_f$ (shell C), and $3.2 \times 10^{16} M_\odot = 3.2 M_f$ (shell D). The dot-dashed line shows the shell trajectory from a calculation without cooling.

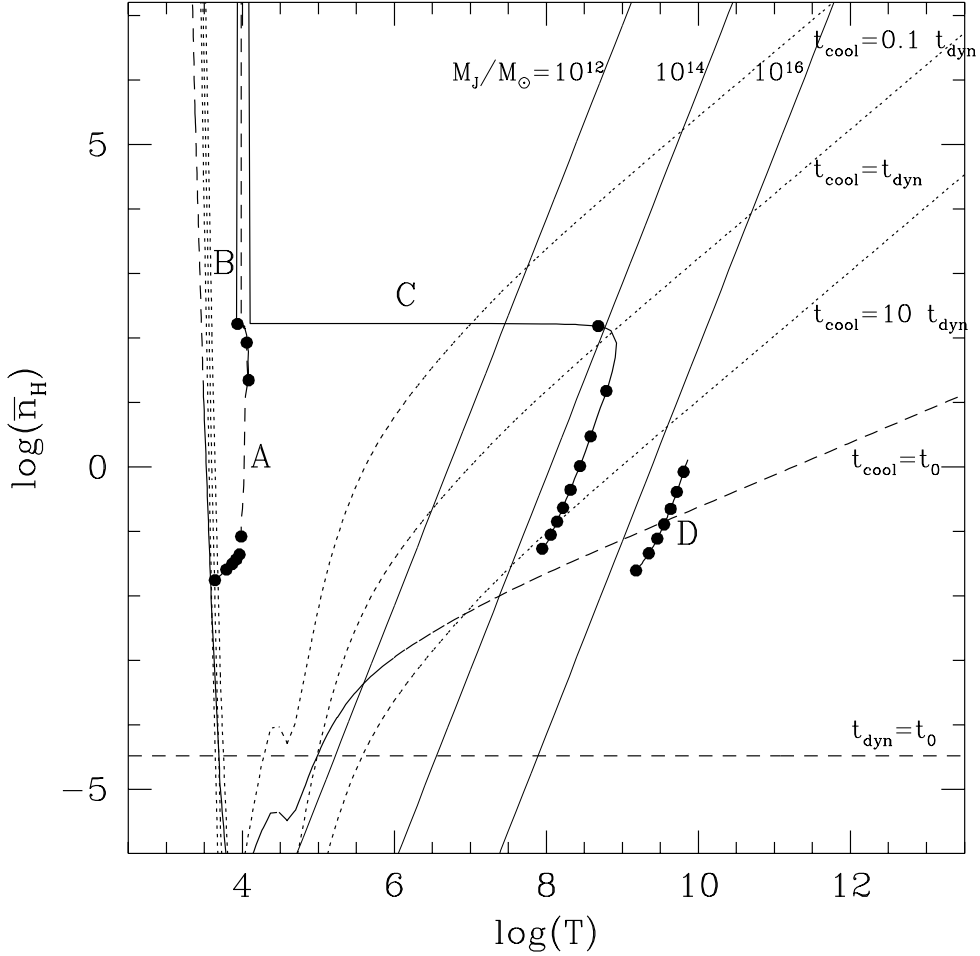

Fig. 9.— Post-shock trajectories in the $\log(T)$-$\log(\bar{n}_H)$ plane of the four fluid shells shown in Figure 8, labeled A-D as before. The trajectory of A is shown by a dashed line, and the trajectories of other shells by solid lines. Points are spaced at intervals $\Delta t = 0.1 t_0$ along trajectory A, $\Delta t = 0.0012 t_0$ along trajectory B, $\Delta t = 0.025 t_0$ along trajectory C, and $\Delta t = 0.1 t_0$ along trajectory D. Dashed lines show the contours $t_{cool} = t_0$ and $t_{dyn} = t_0$. Dotted lines represent contours of constant $t_{cool}/t_{dyn}$. Cooling times are not precise because they depend on local density $n_H$ instead of mean interior density $\bar{n}_H$; we compute them assuming $n_H = 4\bar{n}_H$. Diagonal solid lines are contours of constant Jeans mass $M_J$.

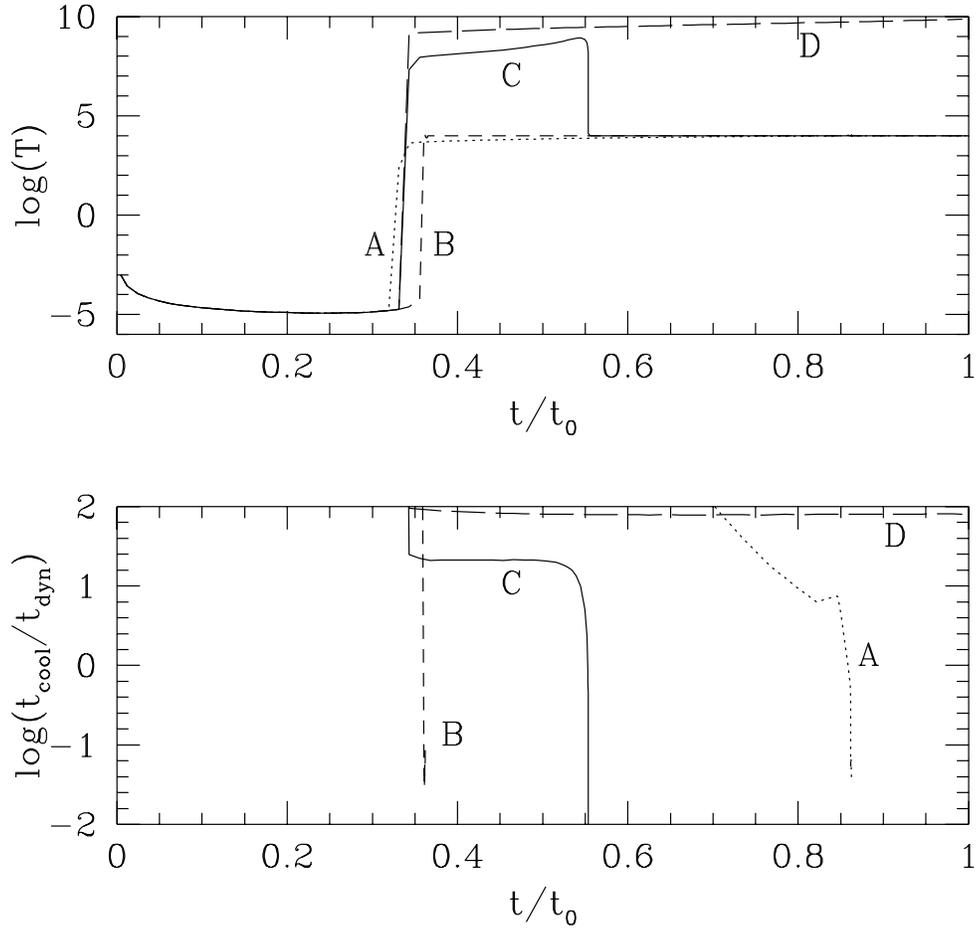

Fig. 10.— Time evolution of the temperature (upper panel) and the ratio $t_{cool}/t_{dyn}$ (lower panel) for the four fluid shells shown in Figures 8 and 9.

Fluid shell B has an interior mass of $3.2 \times 10^8 M_\odot = 3.2 M_f$. For this shell the post-shock temperature is just slightly higher than $10^4$K, in a regime where cooling is very rapid. The shell collapses to $r = 0$ isothermally, at the free-fall rate, with no phase of pressure support. The post-shock evolution is extremely rapid; points along trajectory B in Figure 9 are evenly spaced in time, but the interval is only $\Delta t = 0.0012 \, t_0$.

Shell C has an interior mass of $3.2 \times 10^{14} M_\odot = 3.2 M_f$, six orders of magnitude higher than that of shell B. Immediately after the shock, the cooling time exceeds the dynamical time by more than a factor of 10. Therefore, unlike shell B, shell C goes through a period of nearly adiabatic compression, with little radiative cooling. During this phase, the shell stays in quasi-static equilibrium close to the border of Jeans stability, so it follows a track of constant Jeans mass in the $n - T$ plane (Figure 9). As the density increases, the ratio of cooling time to dynamical time decreases steadily. Once $t_{cool} < t_{dyn}$, the shell cools rapidly to $10^4$K and collapses isothermally at the free-fall rate. Points along trajectory C in Figure 9 are spaced at intervals $\Delta t = 0.025 \, t_0$.

Shell D has an interior mass of $3.2 \times 10^{16} = 3.2 M_f$. The post-shock cooling time is again much longer than a dynamical time. In fact, the cooling time in this case is comparable to the age of the universe, so shell D evolves quasi-statically and never enters a rapid cooling phase. If the simulation were evolved further, it would eventually reach a density high enough for cooling to become important, and it would behave in a fashion more similar to shell C. Points along trajectory D in Figure 9 are spaced at intervals $\Delta t = 0.1 t_0$.

If we want to understand features of the galaxy luminosity function, the quantity of most interest is the mass of gas that actually cools by redshift zero, since this is the gas that could potentially fragment into stars. The filled circles in Figure 11 show the ratio $M_c/M_s$ as a function of $\log(M_s)$, where $M_c$ is the mass of gas that collapses, shocks, and cools by $z = 0$, and $M_s$ is the mass of gas that collapses and shocks by $z = 0$. For $M_s \lesssim 10^{8.5} M_\odot$, the post-shock gas remains below $10^4$K, so it never cools, and $M_c = 0$. Between $M_s = 10^{8.5} M_\odot$ and $M_s = 10^{11} M_\odot$, virtually all of the shocked gas cools, and $M_c/M_s \approx 1$. Above $M_s = 10^{11} M_\odot$, the outermost shocked shells remain pressure supported all the way to $z = 0$ and do not cool. However, the inner shells of these perturbations collapse earlier, at higher density and lower virial temperature, and these shells are able to cool. Therefore, even though the ratio $M_c/M_s$ decreases with increasing $M_s$ in this regime, the decrease is quite slow. In particular, $M_c/M_s$ falls much less rapidly than $M_s^{-1}$, so the actual amount of cooled gas increases with $M_s$. We will return to this important point in §5. The results in Figure 11 are not sensitive to the assumed collapse redshift; we have carried out the same series of calculations for $z_c = 1$, and the trend of $M_c/M_s$ versus $M_s$ is very similar.

### 4.3. Mixed collapses

The basic physical effects described above carry over to the case of a mixed collapse involving gas and collisionless dark matter. Gas shells still collapse and shock, and their subsequent behavior again depends on the cooling time and dynamical time at the post-shock density and temperature. However, in the presence of a mixture of dark matter and gas, the cooling time depends only on the gas parameters, $t_{cool} \sim u\rho_g/\Lambda$, while the dynamical time depends on the total mass, $t_{dyn} \sim \rho_{tot}^{-1}$, where $\rho_{tot} \propto M(r)/r^3$. If we assume that the gas and dark matter density profiles are similar until the point where the gas cools, the ratio between $\rho_g$ and $\rho_{tot}$ is simply given by $\Omega_b/\Omega$. Therefore, we have $t_{cool}/t_{dyn} \propto \rho_{tot}^{-1/2}\Omega/\Omega_b$, and the ratio of the cooling time to the dynamical time is larger than that in the pure fluid case by a factor $\Omega/\Omega_b$. As a consequence, we expect the transition at high masses between the region where all of the gas cools and the region where only part of the gas cools to occur at a lower total mass.

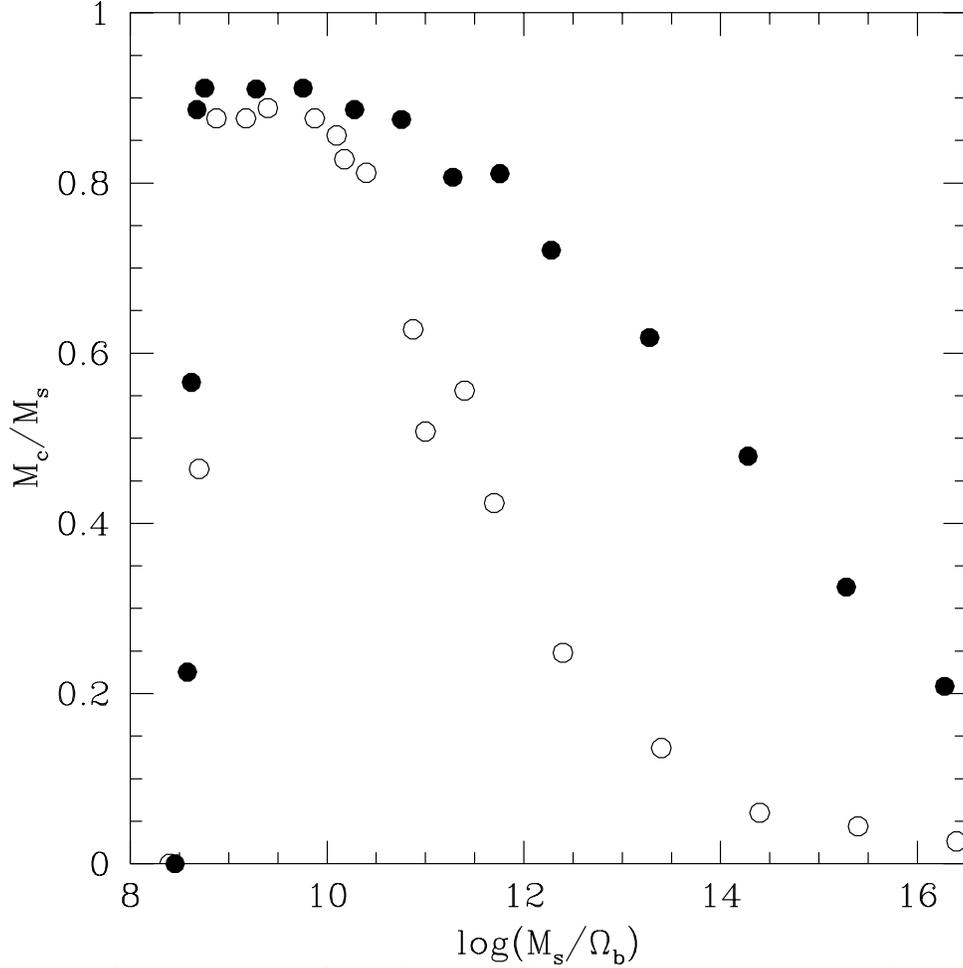

Fig. 11.— Ratio of the mass $M_c$ of gas that cools by $z = 0$ to the mass $M_s$ of gas that shocks by $z = 0$, as a function of the virialized mass $M_s/\Omega_b$. Filled circles represent pure fluid collapses ($\Omega_b = 1$) with collapse redshift $z_c = 2$. Open circles represent mixed collapses with $\Omega_b = 0.1$, $\Omega_d = 0.9$, and $z_c = 2$.

We have again performed a series of collapse calculations in which we vary the value of the filter mass $M_f$ while keeping the collapse redshift fixed at $z_c = 2$. In these calculations, we use $\Omega_d = 0.9$, $\Omega_b = 0.1$, $N_g = 500$ fluid shells, and $N_d = 10,000$ dark matter shells. The open circles in Figure 11 show the results for the ratio of cooled gas to shocked gas, $M_c/M_s$, as a function of $\log(M_s/\Omega_b)$. Once again $M_c$ is the cooled gas mass at $z = 0$ and $M_s$ is the shocked gas mass at $z = 0$. The mass $M_s/\Omega_b$ is, roughly, the total virialized mass at $z = 0$. As expected, the transition at high masses occurs at a lower threshold than in the case of a pure fluid collapse, but the general shape of the curve is similar, and the conclusion is the same, i.e the mass that cools is a monotonically increasing function of the total mass of the perturbation.

## 5. Discussion

Figure 11 shows three distinct regimes: at very low masses, $M_s \lesssim 10^{8.5} M_\odot$, there is no cooling at all, at intermediate masses, $10^{8.5} M_\odot \lesssim M_s \lesssim 10^{11} M_\odot$, all of the shocked gas cools, and at high masses, $M_s \gtrsim 10^{11} M_\odot$, only a fraction of the shocked gas cools. However, the transition from the intermediate-mass regime to the high-mass regime is not a sharp one. Five orders of magnitude above the transition mass, the fraction of shocked gas that is able to cool is still $\sim 20\%$ in the pure fluid case and $\sim 5\%$ in the mixed fluid/dark matter case. We see, therefore, that the requirement that gas be able to cool within a Hubble time cannot by itself explain the sharp upper cutoff in the luminous mass of observed galaxies. This point is further illustrated in Figure 12, where we show the mass of gas that cools by $z = 0$, $M_c$, as a function of the mass of gas that has been shocked, $M_s$. The cooled mass $M_c$ increases monotonically with $M_s$, and the transition between the intermediate- and high-mass regimes is marked only by a modest change of slope in the $M_c$ vs. $M_s$ relation. For large masses the virial temperature is high, the gas is collisionally ionized, and the cooling is dominated by free-free transitions. Including a photo-ionizing background would not, therefore, alter our results at high masses. The low-mass behavior could be sensitive to assumptions about photo-ionization (Efstathiou 1992), a point that we will address elsewhere.

RO suggest that cooling requirements can explain the turnover in the galaxy luminosity function, but the numerical results in Figure 12 indicate otherwise. Larger mass collapses can always produce larger mass galaxies, even if they do so with imperfect efficiency. In essence, the difference between our result and RO's is the difference between a multi-zone and a single-zone calculation. RO associate a single characteristic density and a single characteristic temperature with the collapse of a given perturbation. They then ask whether a cloud of gas at that density and temperature can cool within a Hubble time, or within a dynamical time. However, typical collapses produce peaked density profiles rather than uniform profiles, so the inner, high-density regions can cool more efficiently than the outer regions. This point has been made in a somewhat different guise by White & Frenk (1991), in their semi-analytic models of the galaxy formation process. They assume that a collapse without cooling would produce a cloud with an $r^{-2}$ density profile and a temperature equal to the halo virial temperature. They then compute the "cooling radius" — the radius out to which gas is dense enough to cool within a Hubble time — and from this they compute the cooled mass. This line of reasoning leads to good qualitative agreement with our numerical arguments, as shown by the dashed line in Figure 12, which represents the result of applying the White & Frenk analysis to the case of pure fluid collapses. The transition between complete cooling and incomplete cooling is gradual and subtle, unlike the turnover in the galaxy luminosity function.

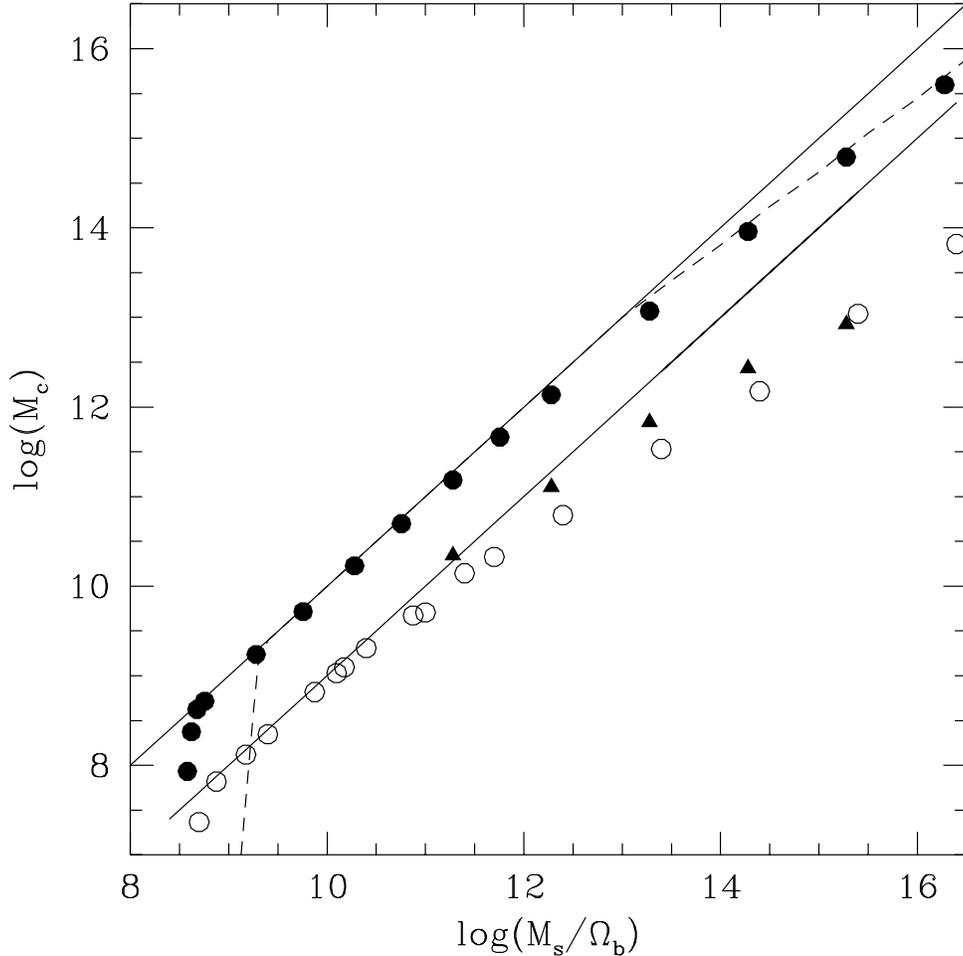

Fig. 12.— The mass $M_c$ of gas that cools by $z = 0$ as a function of the virialized mass $M_s/\Omega_b$. Filled and open circles represent pure fluid and mixed collapses, respectively. Solid lines show $M_c = M_s$, the result that would be obtained for complete cooling of shocked gas. The numerical results fall below these lines at low masses (where there is no cooling) and at high masses (where there is partial cooling), but there is no sharp transition at high masses that might correspond to the turnover in the galaxy luminosity function. The dashed line shows the result of applying the White & Frenk (1991) analysis to the case of pure fluid collapses. Filled triangles show, for the pure fluid case, the mass that cools within a single post-shock dynamical time.

Our spherically symmetric calculation models the process of galaxy formation with a coherent collapse. However, the complex and untidy assembly of a proto-galaxy in a hierarchical scenario still generates a wide range of densities and corresponding cooling times, since material that collapses early reaches higher density and cools more efficiently. We therefore expect that the qualitative trend in Figure 12 would continue to hold in a more realistic calculation. Small-scale clustering in a hierarchical model will also heat gas prior to the proto-galaxy collapse, but the agreement between our numerical results and the White & Frenk analysis (which assumes that all gas is at the virial temperature) implies that it is the range of densities rather than the range of post-shock temperatures that explains the cooling in high-mass collapses. We have explicitly checked this point in the case of a pure fluid, $M_f = 10^{15} M_\odot$ collapse, by running a series of simulations in which the gas is held at a finite temperature prior to shocking. The pre-heating has little or no effect on the mass of gas that cools until the pre-heat temperature reaches $10^{7.9}$K, about 25% of the perturbation's virial temperature. At higher temperatures, pressure support prevents gas shells from turning around and collapsing.

Allowing gas a Hubble time to cool may be unfairly generous for a hierarchical scenario. A perturbation will often merge with another of comparable size only a few dynamical times after it collapses, and gas that has not cooled by then can be shock-heated to a higher temperature. The appropriate time to allow for cooling may therefore be a few dynamical times rather than a Hubble time. The triangles in Figure 12 show, for several pure fluid calculations, the gas that cools within a *single* dynamical time, i.e. a gas shell that shocks at $t_s$ must cool by time $t_s + (3\pi/16\bar\rho)^{1/2}$, where $\bar\rho$ is the average interior density immediately after the shock. Since there is less time for cooling (and the ordinate $M_s$ is still the mass that shocks by $z = 0$), the cooled masses are lower. However, while the trend of $M_c$ vs. $M_s$ is somewhat shallower, it is certainly not flat. Even the stringent requirement of cooling within a single dynamical time cannot by itself produce a sharp cutoff in the galaxy luminosity function.

Radiative cooling is bound to be an important ingredient in the process of galaxy formation. Gas must cool before it can form stars, and dissipation is needed to explain the prevalence of galaxy disks, to explain the difference between the characteristic mass of galaxies and the characteristic mass of rich clusters, and, in most scenarios, to explain the high internal densities of ellipticals and bulges. However, our results indicate that the characteristic mass of galaxies cannot simply be read out of the physical constants that describe gravity and atomic physics. Instead, the form of the galaxy luminosity function must reflect a more subtle interplay between cooling and cosmology, particularly the rates at which perturbations collapse and merge. While this interplay complicates our effort to understand the physics of galaxy formation, it raises the hope that the properties of galaxies may provide important constraints on cosmological parameters, the nature of dark matter, and the spectrum of fluctuations in the early universe.

We thank Simon White for helpful discussions, Martin Rees for comments on the manuscript, and the referee for alerting us to the work of Thomas (1988). This research was supported by the Ambrose Monell Foundation (AT), the W.M. Keck Foundation (DW) and by NSF grant PHY92-45317. We enjoyed the hospitality and stimulating atmosphere of the Aspen Center for Physics during part of this work.

## REFERENCES

Bardeen, J. M., Bond, J. R., Kaiser, N., & Szalay, A. S. 1986, ApJ, 304, 15

Bertschinger, E. 1985, ApJS, 58, 39

Binney, J. J. 1977, ApJ, 215, 483

Black, J. H. 1981, MNRAS, 197, 553

Bowers, R. L. & Wilson, J. R. 1991, Numerical Modeling in Applied Physics and Astrophysics (Boston: Jones and Bartlett Publishers)

Cen, R. 1992, ApJS, 78, 341

Cen, R. & Ostriker, J.P. 1992, ApJ 393, 22

Cen, R. & Ostriker, J.P. 1993, ApJ 417, 404

Cole, S., Aragon-Salamanca, A., Frenk, C. S., Navarro, J. F. & Zepf, S. E. 1994, MNRAS, in press

Evrard, A.E., Summers, F.J & Davis, M. 1994, ApJ 422, 11

Gott, J. R. III 1975, ApJ, 201, 296

Gradshteyn, I. S. & Ryzhik, I. M. 1980, Table of Integrals, Series, and Products (New York: Academic Press)

Katz, N. & Gunn, J. E. 1991, ApJ, 377, 365

Katz, N, Hernquist, L. & Weinberg, D. H. 1992, ApJ, 399, 109

Kauffmann, G., White, S. D. M. & Guiderdoni, B. 1993, MNRAS, 264, 201

Larson, R. B. 1969, MNRAS, 145, 405

Larson, R. B. 1974, MNRAS, 166, 585

Padmanabhan, T. 1993, Structure Formation in the Universe (Cambridge University Press)

Press, W. H. & Schechter, P. 1974, ApJ, 187, 425

Rees, M. J. & Ostriker, J. P. 1977, MNRAS, 179, 541

Richtmyer, R. & Morton, K. W. 1967, Difference Methods for Initial-Value Problems (New York: Interscience Publishers)

Silk, J. I. 1977, ApJ, 211, 638

Shapiro, P. R. & Struck-Marcell, C. 1985, ApJS, 57, 205

Shapiro, S. L. & Teukolsky, S. A. 1983, Black Holes, White Dwarfs, and Neutron Stars, The Physics of Compact Objects (New York: Wiley-Interscience)

Spitzer, L. Jr. & Hart, M. H. 1971, ApJ, 166, 483

Steinmetz, M. & Muller, E. 1994, A&A, 281, L97

Thomas, P. 1988, MNRAS, 235, 315

White, S. D. M. & Rees, M. J. 1978, MNRAS, 183, 341

White, S. D. M. & Frenk, C. S. 1991, ApJ, 379, 52